\documentclass[prodmode,acmcsur]{acmsmall} % Aptara syntax

\usepackage[ruled]{algorithm2e}
\usepackage{subfigure}
\usepackage{bbding}
\usepackage{caption}
\usepackage{longtable}
\usepackage{array}
\usepackage{pdflscape}
\usepackage{rotating}
\usepackage{soul}

\SetAlFnt{\small}
\SetAlCapFnt{\small}
\SetAlCapNameFnt{\small}
\SetAlCapHSkip{0pt}
\IncMargin{-\parindent}

\begin{document}

% Page heads
\markboth{C. Qu et al.}{Auto-scaling Web Applications in Clouds: A Taxonomy and Survey}

% Title portion
\title{Auto-scaling Web Applications in Clouds: A Taxonomy and Survey}
\author{CHENHAO QU
\affil{The University of Melbourne, Australia}
RODRIGO N. CALHEIROS
\affil{Western Sydney University, Australia}
RAJKUMAR BUYYA
\affil{The University of Melbourne, Australia}
}
% NOTE! Affacmsmall-sample-bibfile.bibiliations placed here should be for the institution where the
%       BULK of the research was done. If the author has gone to a new
%       institution, before publication, the (above) affiliation should NOT be changed.
%       The authors 'current' address may be given in the "Author's addresses:" block (below).
%       So for example, Mr. Abdelzaher, the bulk of the research was done at UIUC, and he is
%       currently affiliated with NASA.

\begin{abstract}
Web application providers have been migrating their applications to cloud data centers, attracted by the emerging cloud computing paradigm. One of the appealing features of the cloud is elasticity. It allows cloud users to acquire or release computing resources on-demand, which enables web application providers to automatically scale the resources provisioned to their applications without human intervention under a dynamic workload to minimize resource cost while satisfying Quality of Service (QoS) requirements. In this paper, we comprehensively analyze the challenges that remain in auto-scaling web applications in clouds and review the developments in this field. We present a taxonomy of auto-scalers according to the identified challenges and key properties. We analyze the surveyed works and map them to the taxonomy to identify the weaknesses in this field. Moreover, based on the analysis, we propose new future directions that can be explored in this area.
\end{abstract}




% We no longer use \terms command
%\terms{Design, Algorithms, Performance}


\acmformat{Chenhao Qu, Rodrigo N. Calheiros,
and Rajkumar Buyya, 2016. Auto-scaling Web Applications in Clouds: A Taxonomy and Survey.}
% At a minimum you need to supply the author names, year and a title.
% IMPORTANT:
% Full first names whenever they are known, surname last, followed by a period.
% In the case of two authors, 'and' is placed between them.
% In the case of three or more authors, the serial comma is used, that is, all author names
% except the last one but including the penultimate author's name are followed by a comma,
% and then 'and' is placed before the final author's name.
% If only first and middle initials are known, then each initial
% is followed by a period and they are separated by a space.
% The remaining information (journal title, volume, article number, date, etc.) is 'auto-generated'.

\begin{bottomstuff}
Author's addresses: C. Qu and R. Buyya: Cloud Computing and Distributed Systems (CLOUDS) Laboratory, School of Computing and Information Systems, The University of Melbourne, Australia. R. N. Calheiros: School of Computing, Engineering and Mathematics, Western Sydney University, Australia.
\end{bottomstuff}

\maketitle

\section{Introduction}
\label{sec:introduction}

Cloud computing is the emerging paradigm for offering computing resources and applications as subscription-oriented services on a pay-as-you-go basis. One of its features, called elasticity, allows users to dynamically acquire and release the right amount of computing resources according to their needs. Elasticity is continuously attracting web application providers to move their applications into clouds.

To efficiently utilize elasticity of clouds, it is vital to automatically and timely provision and deprovision cloud resources without human intervention, since over-provisioning leads to resource wastage and extra monetary cost, while under-provisioning causes performance degradation and violation of service level agreement (SLA). This mechanism of dynamically acquiring or releasing resources to meet QoS requirements is called auto-scaling. 

Designing and implementing an efficient general-purpose auto-scaler for web applications is a challenging task due to various factors, such as dynamic workload characteristics, diverse application resource requirements, and complex cloud resources and pricing models. In this paper, we aim to comprehensively analyze the challenges in the implementation of an auto-scaler in clouds and review the developments for researchers that are new to this field. We present a taxonomy regarding the various challenges and key properties of auto-scaling web applications. We compare the existing works deployed by infrastructure providers and application service providers and map them to the taxonomy to discuss their strengths and weaknesses. Based on the analysis, we also propose promising future directions that can be pursued by researchers to improve the state-of-the-art.

\cite{Lorido-Botran2014} have already written a survey about this topic. However, their major focus is on resource estimation techniques. Different from them, our work provides comprehensive discussions about all the primary challenges in the topic and it also introduces new developments that occurred since their work was published. %We first define the auto-scaling problem for web applications in clouds and then identify the key challenges that need to be addressed when trying to implement one. We propose a taxonomy of the existing auto-scaling systems based on the methods employed to tackle the various challenges. Follow the taxonomy, we describe and compare the important works that have been done in this field. Finally, we give our thoughts of potential future directions to push the state-of-the-art further.

The rest of the paper is organized as follows. In Section~\ref{sec:problem_definition}, we describe our definition of the auto-scaling problem for web applications and list its major challenges that need to be addressed when trying to implement an auto-scaler. After that, we present a taxonomy regarding the existing auto-scalers. From Section~\ref{sec:detail_start} to Section~\ref{sec:detail_end}, we introduce and compare how existing auto-scaling techniques tackle the listed challenges. After that, in Section~\ref{sec:discussion}, we discuss the gaps of current solutions and present some promising future research directions. Finally, we summarize the findings and conclude the paper.

\section{Problem Definition and Challenges}
\label{sec:problem_definition}

\begin{figure*}[!t]
\begin{center}
\subfigure[Scaling Out/Up]
{\label{fig:scaling_up}
\includegraphics[width=5.6in]{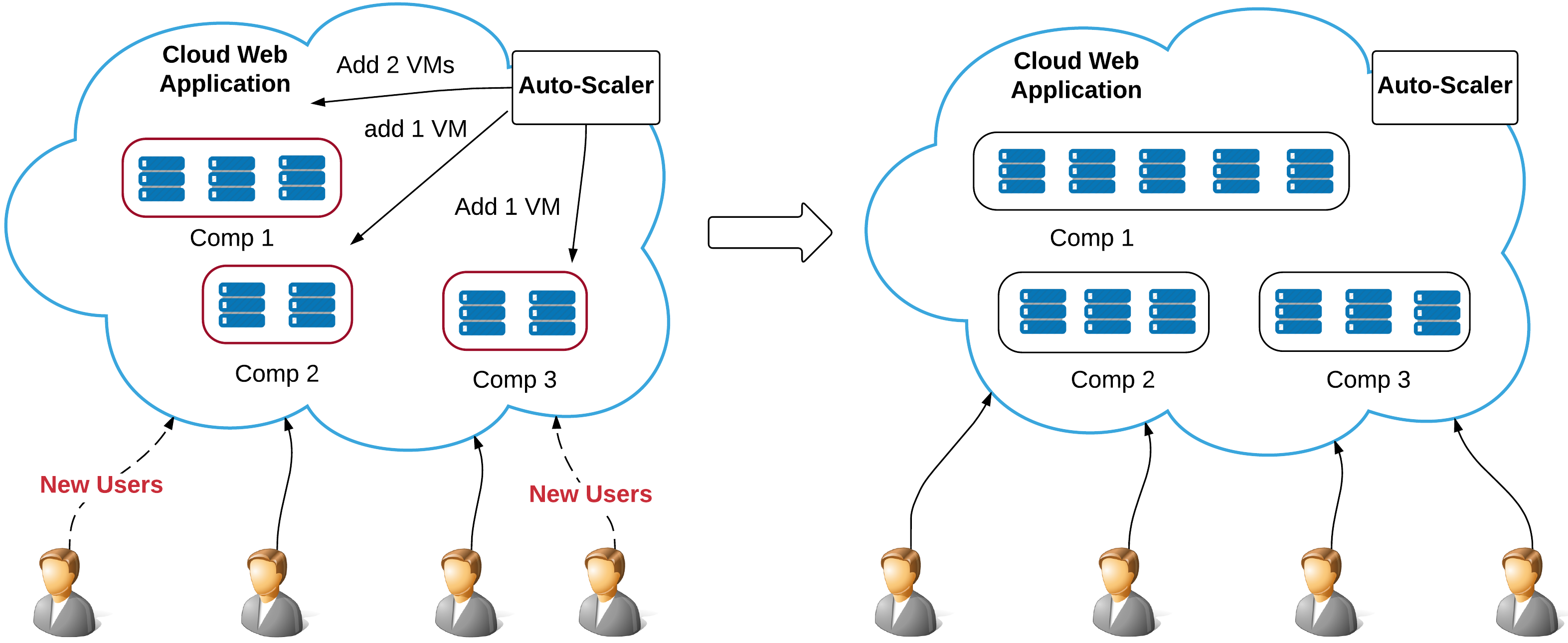}}
\subfigure[Scaling In/Down]
{\label{fig:scaling_down}
\includegraphics[width=5.6in]{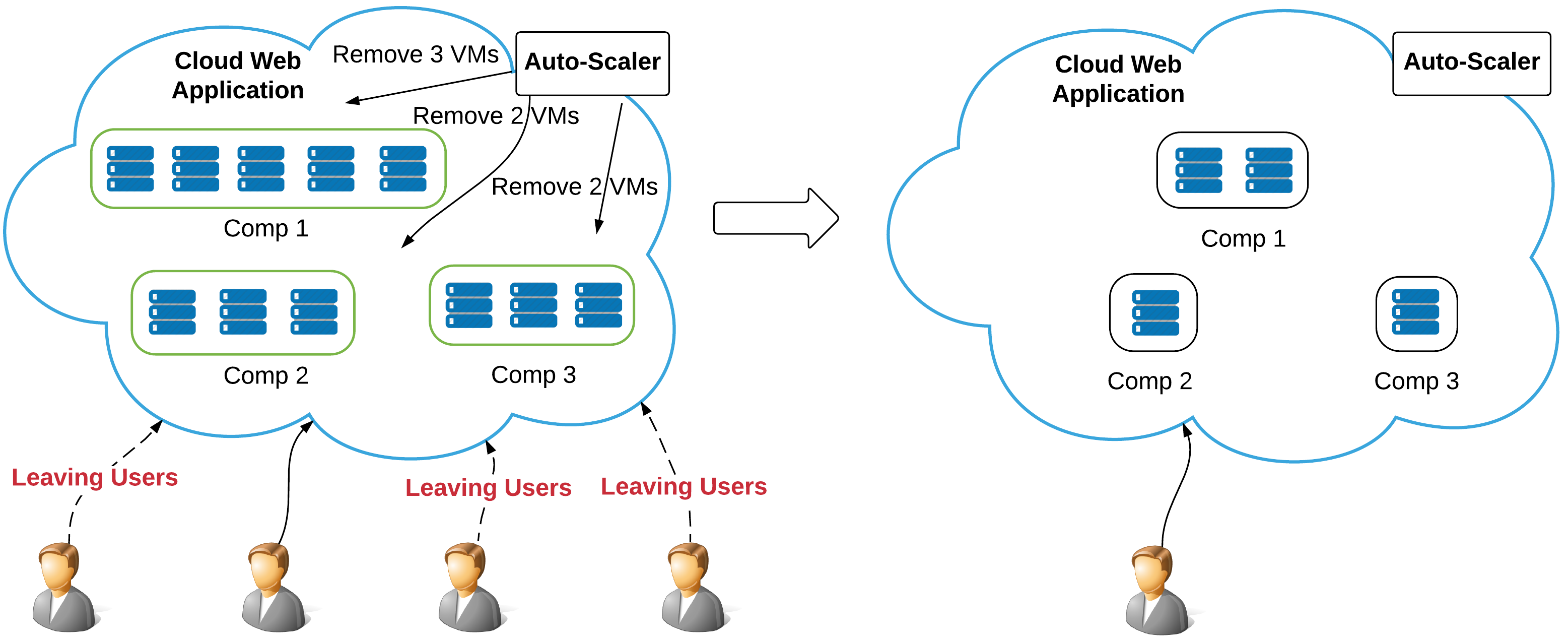}}
\end{center}
\caption{Typical auto-scaling scenarios --- right sizing of resources}
\label{fig:auto-scaling}
\end{figure*}

\begin{figure}
\centering
\includegraphics[width=2.5in]{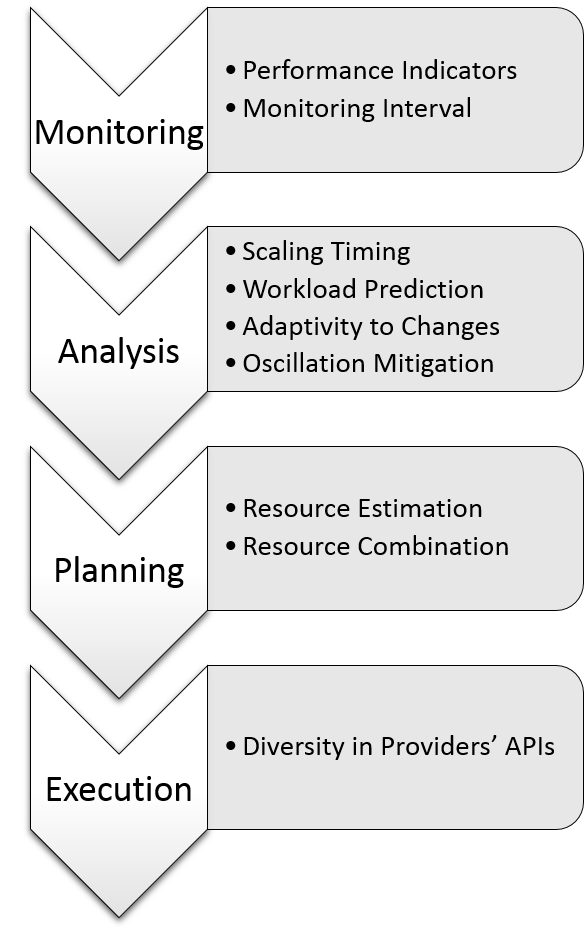}
\caption{The challenges of auto-scaling web applications in each phase of the MAPE loop}
\label{fig:challenges}
\end{figure}

In a single cloud, the auto-scaling problem for web applications can be defined as how to autonomously and dynamically provision and deprovision a set of resources to cater to fluctuant application workloads without human intervention so that the resource cost is minimized and application service level agreements (SLAs) or service level objectives (SLOs) are satisfied. Figure~\ref{fig:auto-scaling} illustrates typical auto-scaling scenarios. In Figure~\ref{fig:scaling_up}, due to increase in requests, the available resources are in congestion, and thus, the auto-scaler decides to provision certain resources by either scaling out (launching more VMs) or scaling up (adding resources to existing VMs) each application component. Oppositely, in Figure \ref{fig:scaling_down}, the auto-scaler deprovisions some resources from each component by either scaling in (shutting down some VMs) or scaling down (removing resources from existing VMs), when the amount of requests has decreased.

This is a classic automatic control problem, which demands a controller that dynamically tunes the type of resources and the amount of resources allocated to reach certain performance goals, reflected as the SLA. Specifically, it is commonly abstracted as a MAPE (Monitoring, Analysis, Planning, and Execution) control loop~\cite{Kephart2003}. The control cycle continuously repeats itself over time.

The biggest challenges of the problem lie in each phase of the loop as shown in Figure \ref{fig:challenges}. We briefly explain each phase and summarize the individual challenges faced by auto-scaler designers in the following paragraphs.

\paragraph{Monitoring}
Auto-scaler needs to monitor some performance indicators to determine whether scaling operations are necessary and how they should be performed.

\begin{itemize}
\item Performance indicators: selection of the right performance indicators is vital to the success of an auto-scaler. The decision is often affected by many factors, such as application characteristics, monitoring cost, SLA, and the control algorithm itself.

\item Monitoring interval: monitoring interval determines the sensitivity of an auto-scaler. However, very short monitoring intervals result in high monitoring cost both regarding computing resources and financial cost, and it is likely to cause oscillations in the auto-scaler. Therefore, it is important to tune this parameter to achieve balanced performance.
\end{itemize}

\paragraph{Analysis}
During the analysis phase, the auto-scaler determines whether it is necessary to perform scaling actions based on the monitored information.

\begin{itemize}
\item Scaling timing: the auto-scaler firstly needs to decide when to perform the scaling actions. It can either proactively provision/deprovision resources ahead of the workload changes if they are predictable since the provision/deprovision process takes considerable time or it can perform actions reactively when workload change has already happened.

\item Workload prediction: if the auto-scaler chooses to scale the application proactively, how to accurately predict the future workload is a challenging task.

\item Adaptivity to changes: sometimes the workload and the application may undergo substantial changes. The auto-scaler should be aware of the changes and timely adapt its model and settings to the new situation.

\item Oscillation mitigation: scaling oscillation means the auto-scaler frequently performs opposite actions within a short period (i.e., acquiring resources and then releasing resources or vice versa). This situation should be prevented as it results in resource wastage and more SLA violations.
\end{itemize}

\paragraph{Planning}
The planning phase estimates how many resources in total should be provisioned/deprovisioned in the next scaling action. It should also optimize the composition of resources to minimize financial cost.

\begin{itemize}

\item Resource estimation: the planning phase should be able to estimate how many resources are just enough to handle the current or incoming workload. This is a difficult task as the auto-scaler needs to determine required resources quickly without being able to actually execute the scaling plan to observe the real application performance, and it has to take the specific application deployment model into account in this process.

\item Resource combination: to provision resources, the auto-scaler can resort to both vertical scaling and horizontal scaling. If horizontal scaling is employed, as the cloud providers offer various types of VMs, the auto-scaler should choose one of them to host the application. Another important factor is the pricing model of cloud resources. Whether to utilize on-demand, reserved or rebated resources significantly affects the total resource cost. All these factors form a huge optimization space, which is challenging to be solved efficiently in short time.

\end{itemize}

\paragraph{Execution}
The execution phase is responsible for actually executing the scaling plan to provision/deprovision the resources. It is straightforward and can be implemented by calling cloud providers' APIs. However, from an engineering point of view, being able to support APIs of different providers is a challenging task.

\begin{sidewaysfigure}
\centering
\includegraphics[width=8in]{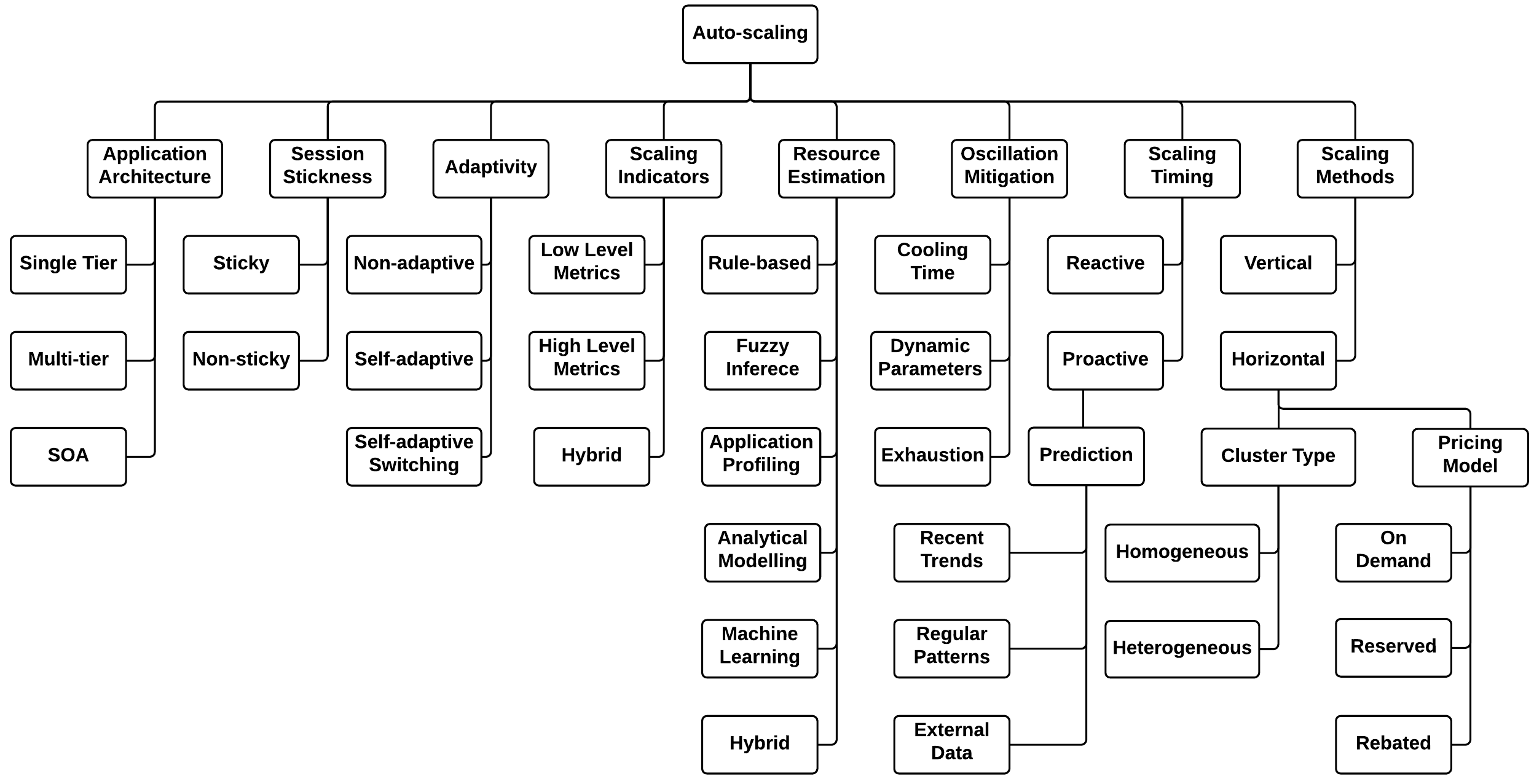}
\caption{The taxonomy for auto-scaling web applications in clouds}
\label{fig:taxonomy}
\end{sidewaysfigure}

If the application is supposed to be deployed in multiple data centers, it is also important to identify which data center is most cost-efficient to serve the requests from certain groups of users without violating SLAs. Therefore, in addition to provisioning just enough amount of resources during runtime, the auto-scaling problem becomes a mixed problem of data center selection, geographical load balancing, and resource provisioning in multi-cloud scenario. The auto-scaler should dynamically direct users from certain areas to specific data centers, and ensure enough resources are provisioned in each of the involving data centers to handle the incoming requests. To minimize cost in this scenario, considering all available choices in these tasks, it generally requires solving a NP-hard problem to generate the provision plan.

\section{Taxonomy}
\label{sec:taxonomy}

Figure \ref{fig:taxonomy} illustrates our proposed taxonomy for auto-scaling web applications in clouds. It classifies the existing works based on the identified challenges in each of the MAPE phase in section~\ref{sec:problem_definition} and their targeted environment. Particularly, the taxonomy covers the following aspects of auto-scaling:

\begin{itemize}
\item Application Architecture: the architecture of the web application that the auto-scaler is managing.

\item Session Stickiness: whether the auto-scaler supports sticky sessions.

\item Adaptivity: whether and how the auto-scaler adapts to changes of workload and application.

\item Scaling Indicators: what metrics are monitored and measured to make scaling decisions.

\item Resource Estimation: how the auto-scaler estimates the amount of resources needed to handle the workload.

\item Oscillation Mitigation: how the auto-scaler reduces the chance of provisioning oscillation.

\item Scaling Timing: whether the auto-scaler supports proactive scaling of applications and how it predicts future workload.

\item Scaling Methods: how the auto-scaler decides the methods to provision resources and what combination of resources are provisioned to the application.

\item Environment: whether the auto-scaler works in a single or multi-cloud environment. 
\end{itemize}

Note that the considered characteristics of auto-scaling may be correlated, especially for resource estimation and oscillation mitigation. Resource estimation is affected by many other characteristics, such as application architecture, adaptivity, and scaling indicators. Oscillation mitigation is often influenced by the choice of resource estimation model and the scaling methods used. An existing approach generally spans across different subcategories and are discussed in each of them (i.e., an auto-scaler is built in a single cloud environment for multi-tier applications, and employs proactive scaling with machine learning resource estimation techniques). Note that this taxonomy is based on features and thus does not reflect the relative performance of the proposed approaches. Actually, because the surveyed works target diverse workload patterns, application architectures, and pricing models, there is no single answer to the question of which approach generally performs the best \cite{Papadopoulos2016}.

In the following sections (from Section~\ref{sec:detail_start} to Section~\ref{sec:detail_end}), we introduce and compare existing auto-scalers according to this taxonomy.

\section{Application Architectures}
\label{sec:detail_start}

There are three types of web application architectures mentioned in the literature: namely single tier, multi-tier, and service-oriented architecture.

\subsection{Single Tier/Single Service}
A tier is the minimum separately deployable component in a layered software stack. In a production deployment, within a tier, a load balancer is used to balance and dispatch load among the instances of the same tier. Single tier architecture by definition is the architecture in which an application is composed of only one tier. Relatively, the architecture with multiple connected software tiers is called multi-tier architecture. Instead of calling single tier as an application architecture, it is more accurate to think of it as the smallest granularity that can be possibly managed by an auto-scaler, since hardly any web application is composed of only one tier.

Nowadays, web applications are becoming more and more complicated and deviate from the traditional multi-tier architecture. In those cases, the fundamental scaling component is often referred as a service/microservice. The majority of existing auto-scalers separately manage each single tier or service within an application instead of considering it as a whole. This method is both simple and general. However, it often results in globally suboptimal resource provisioning as it requires to divide the SLA requirements of the overall application into sub-requirements of each tier or service, which is often a challenging and subjective task.

%\begin{table}[!t]
%\tbl{Auto-scaling systems for Multi-tier applications in the literature}{
%\begin{tabular} {c | c}
%\hline
%\hline
% \textbf{Work} & \textbf{SLA}\\
%\hline
 
% Gandhi \cite{Gandhi} & DAC\\
%\hline
%\end{tabular}}
%\end{table}

\subsection{Multi-tier}
\label{subsec:multi-tier}
Multi-tier applications, as introduced, are composed of sequentially connected tiers. At each tier, the request either relies on the downstream tier to complete its processing or it is returned to the upstream tier and finally to the user.

A widely-adopted architecture of this type usually consists of three tiers: one frontend, one application logic, and one database tier. The database tier is often considered dynamically unscalable and ignored by auto-scalers.

Many works have targeted multi-tier applications. Some of them employ the divide and conquer approach that breaks overall SLA into SLA of each tier, such as the works conducted by \cite{Urgaonkar2008}, \cite{Singh2010},  \cite{Iqbal2011}, \cite{Malkowski2011}, \cite{Upendra}, and \cite{Gergin}. Others consider SLA of the whole application and provision resources to each tier holistically. This strategy requires more efforts in modeling and estimating resource consumption using sophisticated queuing networks and machine learning techniques as discussed in Section~\ref{sec:resource_estimation}, and the resulting auto-scalers are only applicable to multi-tier applications. Important works of this category include approaches proposed by \cite{Zhang}, \cite{Jung}, \cite{Padala2009}, \cite{Lamaa,Lama2010}, \cite{Sharma2012}, \cite{Han2014}, and \cite{Kaur2014}.

\subsection{Service-based Architectures}
\label{subsec:SOA}
Service-based architectures have now become the dominant paradigm for large web applications, such as Amazon e-commerce website and Facebook. In these architectures, applications are composed of standalone services that interact with each other through pre-defined APIs. More importantly, services are not necessarily connected sequentially as in multi-tier applications. Service-based applications are commonly abstracted as directed graphs with each nodes representing services and directed edges representing their interactions. The service-based architectures can be further classified into Service-oriented Architecture (SOA) and Microservices Architecture. \cite{Richards2015} provides a full discussion of their commonalities and differences, which is out of the scope of this paper. From the perspective of auto-scaling, they pose the same question as how to provision individual services so that the QoS requirements of the aggregated application can be satisfied.

Due to the complexity of service-based architectures, it is difficult to manage resource provisioning of all the services holistically. Therefore, industry and most works employ the divide and conquer approach. Differently, \cite{Jianga} proposed a method that can satisfy SLA of the whole SOA application. It requires each service to estimate the change of its response time if one instance is added or removed from it. After that, the system aggregates the estimations and chooses the operations that will minimize the overall response time.

\section{Session Stickiness}

A session is a series of interactions between a client and the application. After each operation, the client waits for the reply given by the application and then proceeds. To ensure a seamless experience, it is necessary to keep the intermediate statuses of clients during their sessions. Otherwise, the operations conducted by the clients will be lost and they have to repeat the previous operations to proceed. Taking a social network application as an example, a session may involve the following operations: the client first accesses the home page and then logs into the application; after that, he performs several actions such as viewing his and his friends' timeline, uploading photos, and updating his status, before he quits the application.

This session based access pattern has caused issues on efficiently utilizing elastic resources in cloud because the stateful nature of session forces the user to be connected to the same server each time he submits a request within the session if the session data is stored in the server. Such sessions are considered sticky. They limit the ability of the auto-scaler to terminate under-utilized instances when there are still unfinished sessions handled by them. Therefore, it is regarded a prerequisite to transforming stateful servers into stateless servers before an auto-scaler can manage them.

There are multiple ways to achieve this, and a complete introduction to them is out of the scope of this paper. The most adopted approach is to move the session data out of the web servers and store them either at user side or in a shared Memcached cluster.

Though most auto-scalers require the scaling cluster to be stateless, some auto-scalers are designed to handle stateful instances. \cite{Chieua} proposed an auto-scaler based on the number of active sessions in each server. Their approach requires a server to clear all its active sessions before it can be terminated. \cite{Grozev2014} proposed a better approach by integrating a similar auto-scaler with a load balancing algorithm that consolidates sessions within as few instances as possible.

\section{Adaptivity}
Auto-scalers fall in the realm of control systems. As stated in the introduction, their operation involves tuning resources provisioned to the application to reach the target performance. One major issue coupled with the design of a control system is its adaptivity to changes. As in dynamic production environments, workload characteristic, and even the application itself can change at any moment. Therefore, adaptivity is important to auto-scalers. Based on the level of adaptivity, we classify the existing works into three categories.

\subsection{Non-adaptive}
In the non-adaptive approaches, the control model is predefined, and they make decisions purely based on the current input. Examples are the rule-based approaches employed by the industry, such as Amazon Auto-Scaling service \cite{Amazon2016}. They require the user to define a set of scaling out and scaling in conditions and actions offline. During production time, the auto-scaler makes scaling decisions only when the conditions are met. They do not allow automatic adjustment of the settings during production. When using this kind of auto-scalers, the users often need to direct considerable efforts in offline testing to find a suitable configuration.

\subsection{Self-adaptive}
Self-adaptive auto-scalers are superior to their non-adaptive counterparts. Though the core control models in them are fixed as well, they are capable of autonomously tuning themselves according to real-time quality of the control actions observed. In this way, the designer only needs to determine the core control model, such as whether it is linear or quadratic, and the auto-scaler will adjust and evolve itself to meet the target performance. This feature can be implemented through extending existing self-adaptive control frameworks in control theory, such as \cite{Kamra2004}, \cite{Kalyvianaki2009}, and \cite{Grimaldi2015}. Self-adaptivity can also be realized through dynamic measurement or correction of parameters in analytical models and machine learning approaches, such as reinforcement learning and regression. Detailed explanations of them are given in Section~\ref{sec:resource_estimation}.

The benefit of introducing self-adaptivity is that it significantly reduces the amount of offline preparation required to utilize an auto-scaler. Furthermore, once substantial changes are detected, self-adaptive approaches can autonomously abort the current model and retrain itself, thus reducing maintenance efforts. Their primary drawback is that it usually takes time for them to converge to a good model and the application will suffer from bad performance during the early stage of training.

\subsection{Self-adaptive Switching}
Beyond the utilization of a single self-adaptive module, some auto-scalers have employed a more adaptive framework, which we call self-adaptive switching. In these auto-scalers, they concurrently connect multiple non-adaptive or self-adaptive controllers and actively switch control between controllers based on their observed performance on the application. The included self-adaptive controllers continuously tune themselves in parallel. However, at each moment, only the selected best controller can provision resources. \cite {Patikirikorala} employed this approach and \cite{Ali-Eldin2013} proposed a self-adaptive switching approach based on the classification of the application workload characteristics, i.e., their periodicity and the burstiness. \cite{Chen2016} proposed an approach that trains multiple resource estimation models and dynamically selects the one that performs the best.

\section{Scaling Indicators}
\label{sec:scaling_indicators}

The actions of auto-scalers are based on performance indicators of the application obtained in the monitoring phase. These indicators are produced and monitored at different levels of the system hierarchy from low-level metrics at the physical/hypervisor level to high-level metrics at the application level.

\subsection{Low-Level Metrics}
Low-level metrics, in the context of this survey, are server information monitored at the physical server/virtual machine layer by hypervisors, such as utilization of CPU, memory, and network resources, memory swap, and cache miss rate. These data can be obtained through a monitoring platform of the cloud provider or from monitoring tools for operating systems. However, it is a non-trivial task to accurately infer the observed application performance merely according to the low-level metrics, and therefore, it makes it a difficult task to make sure that the SLA can be met faithfully with the available resources.

Designing an auto-scaler solely based on low-level performance indicators is possible. The simplest solution is to use the utilization of CPU and other physical resources as indicators and scale resources horizontally or vertically to maintain the overall utilization within a predefined upper and lower bound. Industry systems widely adopt this approach. If the auto-scaler of this kind only supports horizontal scaling, it can be utilized by both cloud providers and service providers. Otherwise, only cloud providers can deploy it as service providers lack the control over underlying resource allocation in the current cloud market.

\subsection{High-Level Metrics}
\label{subsec:high_level_metrics}
High-level metrics are performance indicators observed at the application layer. Only auto-scalers deployed by service providers can utilize these metrics as they are not visible to cloud providers. Useful metrics to auto-scaling include request rate, average response time, session creation rate, throughput, service time, and request mix.

Some metrics, such as request rate, average response time, throughput, and session creation rate, are easy to measure. They alone enable operation of an auto-scaler as they are direct indicators of the performance of the applications and can be used to trigger scaling operations. The easiest method to construct one is to replace utilization metrics in the simple auto-scaler mentioned in the previous section with any of such high-level metrics. However, this approach is not able to accurately estimate the amount of resources needed and often over or under-provision resources.

Some auto-scalers require obtaining information about request service time and request mix \cite{Zhang,Singh2010,Kaur2014} to estimate how much resources are needed. Different from the previous metrics, they cannot directly trigger scaling actions but can be used to support making efficient scaling plans. In addition, these metrics are not straightforward to measure.

Service time is the time a server spends on processing the request, which is widely used in the queuing models to approximate the average response time or sojourn time. Except for a few works \cite{Gergin,Han2014} that assume this metric as known a priori, to accurately measure it, either offline profiling \cite{Prodan2009} or support from the application \cite{Aniello2014} is required. Therefore, instead of directly probing it, some works use other approaches to approximate it. \cite{Kaur2014} mentioned the use of past server logs to infer the mean service time. \cite{Gandhi} employed Kalman filters to estimate service time during runtime. \cite{Zhang} used a regression method to make the approximation. \cite{Jiang2011} resorted to profiling each server when it is first online using a small workload without concurrency and then estimating service time through queuing theory. In another work, \cite{Jianga} utilized a feedback control loop to adjust the estimation of service time at runtime.

Request mix is hard to measure because an understanding of the application is essential to distinguish different types of requests. Designing a mechanism to accurately classify various types of requests from outside of the application itself is an interesting and challenging problem to be explored.

\subsection{Hybrid Metrics}
In some auto-scalers, both high-level and low-level metrics are monitored. A common combination is to observe request rate, response time, and utilization of resources. Some works \cite{Prodan2009,Gandhi} monitor them because the queuing models employed for resource estimation require them as input. Some \cite{Jing2007,Padala2009,Dutta,Yazdanov2013,Fernandez2014} use these hybrid metrics to dynamically build a model relating specific application performance to physical resource usage through online profiling, statistical, and machine learning approaches, thus increasing the accuracy of resource estimation without constructing complex analytical models. Another important reason to monitor request rate along with low-level metrics is to conduct future workload prediction \cite{Roy,Dutta}.

Besides low-level and high-level metrics from the platform and application, other factors outside may also play a significant role. For example, \cite{Frey2014} in their fuzzy-based approach utilizes other related data, such as weather, and political events to predict workload intensity.

\section{Resource Estimation}
\label{sec:resource_estimation}
Resource estimation lies in the core of auto-scaling as it determines the efficiency of resource provisioning. It aims to identify the minimum amount of computing resources required to process a workload to determine whether and how to perform scaling operations. Accurate resource estimation allows the auto-scaler to quickly converge to the optimal resource provisioning. On the other hand, estimation errors either result in insufficient provisioning, which leads to prolonged provisioning process and increased SLA violations, or over provisioning, which incurs more cost.

Various attempts have been made to develop resource estimation models from basic approaches to methods with sophisticated models. We categorize them into six groups, namely rule-based, fuzzy inference, application profiling, analytical modeling, machine learning, and hybrid approaches. In the following subsections, we explain and compare the existing approaches in each group.

\subsection{Rule-based Approaches}
Rule-based approaches are widely adopted by industry auto-scalers, such as Amazon Auto-Scaling Service \cite{Amazon2016}. Its kernel is a set of predefined rules consisting of triggering conditions and corresponding actions, such as ``If CPU utilization reaches 70\%, add two instances'', and ``If CPU utilization decreases below 40\%, remove one instance''. As stated in section \ref{sec:scaling_indicators}, users can use any metrics, low-level or high-level, to define the triggering conditions, and the control target of the auto-scaler is usually to maintain the concerned parameters within the predefined upper and lower threshold. Theoretically, simple rule-based approaches involves no accurate resource estimation; only empirical estimation, which is hardcoded in the action part of the rule as adding or removing certain amount or percentage of instances. As the simplest version of auto-scaling, it commonly serves as benchmark for comparison and is used as the basic scaling framework for works that focus on other aspects of auto-scaling, such as \cite{Dawoud2012}, which aims to compare vertical scaling and horizontal scaling, and \cite{Rui2012}, which considers all possible scaling methods, or prototyping works, like the one carried out by \cite{Iqbal2009}.

Though simple rule-based auto-scaler is easy to implement, it has two significant drawbacks. The first is that it requires understanding of the application characteristics and expert knowledge to determine the thresholds and proper actions. \cite{Al-Haidari2013} conducted a study to show that these parameters significantly affect the auto-scaler's performance. The second is that it cannot adapt itself when dynamic changes occur to the workload or to the application.

Hardcoded number of instances to scale out and scale in, called step sizes, would become inappropriate when the workload changes dramatically. For example, if the application is provisioned by four instances at the start, adding one instance will boost 25\% of the capability. After a while, the cluster has increased to ten instances due to workload surge, adding one instance in this case only increases 10\% of capacity. Improvements are made to the basic model using adaptive step sizes. \cite{Netto2014} proposed an approach that decides the step size holistically at runtime based on the upper threshold, the lower threshold, and the current utilization. It first deduces the upper and lower bounds respectively for step sizes of scaling out and scaling in operations to prevent oscillation and then scale the step sizes using a fixed parameter representing aggressiveness of the auto-scaler determined by the user. They reported the adaptive strategy performed best for bursty workload but led to limited improvements for other types of workloads. \cite{Cunha2014} employed a similar approach. However, in their approach, the aggressiveness parameter is also dynamically tunable according to QoS requirements.

In addition to the step size, fixed thresholds also could cause inefficient resource utilization. For instance, the thresholds of 70\% and 40\% may be suitable for a small number of instances but are inefficient for large clusters as single instance has a subtle impact on the overall utilization and a lot of instances actually can be removed before the overall usage reaching the 40\% lower bound. A solution to mitigate this problem is also to make the thresholds dynamic. \cite{Lim2009,Lim2010} used this approach.

\cite{RightScale2016} proposes another important variation of the simple rule-based approach. Its core idea is to let each instance decide whether to shrink or expand the cluster according to predefined rules and then utilize a majority voting approach to make the final decision. \cite{Calcavecchia2012} also proposed a decentralized rule-based approach. In their proposal, instances are connected as a P2P network. Each instance contacts its neighbors for their statuses and decides whether to remove itself or start a new instance in a particular probability derived from their statuses.

\subsection{Fuzzy Inference}

Fuzzy-based auto-scalers can be considered as advanced rule-based approaches as they rely on fuzzy inference, the core of which is a set of pre-defined If-Else rules, to make provisioning decisions. The major advantage of fuzzy inference compared to simple rule-based reasoning is that it allows users to use linguistic terms like ``high, medium, low'', instead of accurate numbers to define the conditions and actions, which makes it easier for human beings to effectively represent their knowledge (human expertise) about the target. Fuzzy inference works as follows: the inputs are first fuzzifized using defined membership functions; then the fuzzified inputs are used to trigger the action parts in all the rules in parallel; the results of the rules are then combined and finally defuzzified as the output for control decisions. Representative approaches of this kind include the one proposed by \cite{Frey2014} and the work conducted by \cite{Lamaa}. Since manually designing the rule set is cumbersome and manual rule sets cannot timely handle environmental and application changes, fuzzy-based auto-scalers are commonly coupled with machine learning techniques to automatically and dynamically learn the rule set \cite{Jing2007,Jamshidi2016,Lama2010}. Their details are introduced in Section~\ref{subsec:hybrid_approaches}.

\subsection{Application Profiling}
We define profiling as a process to test the saturation point of resources when running the specific application using synthetic or recorded real workload. Application profiling is the simplest way to accurately acquire the knowledge of how many resources are just enough to handle the given workload intensity concurrently. Tests need to be conducted either offline or on the fly to profile an application.

Offline profiling can produce the complete spectrum of resource consumption under different levels of workload. With the obtained model, the auto-scaler can more precisely supervise the resource provisioning process. \cite{Upendra}, \cite{Gandhi2012}, \cite{Fernandez2014}, and \cite{Qu2016} employed this approach. The drawback of it is that the profiling needs to be reconducted manually every time the application is updated.

Profiling can be carried out online to overcome this issue. However, the online environment prohibits the auto-scaler to fine-grainedly profile the application as a VM should be put into service as soon as possible to cater to the increasing workload. \cite{Vasic2012} proposed an approach that first profiles the application, then classifies the application signatures into different workload classes (number of machines needed). When changes happen to the application, the profiled new application characteristics are fed into the trained decision tree to realize quick resource provisioning by finding the closest resource allocation plan stored before. \cite{Nguyen2013} relied on online profiling to derive a resource estimation model for each application tier. When profiling each tier, other tiers are provisioned with ample resources. In this way, one by one, models for all the tiers are obtained. \cite{Jiang2011} proposed a quick online profiling technique for multi-tier applications by studying the correlation of resource requirements that different tiers pose on the same type of VM and the profile of a particular tier on that type of VM. This approach allows them to roughly deduce performance of the VM on each tier without actually running each tier on it. Thus, the newly acquired VM can be put into service in relatively quicker speed.

\subsection{Analytical Modeling}
Analytical modeling is a process of constructing mathematical models based on theory and analysis. For resource estimation problems in auto-scaling, the predominant models are built upon queuing theory.

In the generalized form, a queue can be represented as \emph{A}/\emph{S}/\emph{C}, where \emph{A} is the distribution of time interval between arrivals to the queue, \emph{S} is the distribution of time required to process the job, and \emph{C} stands for the number of servers. Common choices for \emph{A} in the existing works are M (Markov) which means that arrivals follow the Poisson process, and G (General) which means the inter-arrival time has a general distribution. For \emph{S}, the prominent alternatives are M (Markov) which represents exponentially distributed service time, D (Deterministic) which means the service time is fixed, and G (General) which stands the service time has a general distribution. Detailed introduction of different types of queues is out of the scope of this paper. Interested users can refer to \cite{Gnedenko1989}. 

For a single application, tier, or service, if the underlying servers are homogeneous, it is more convenient to abstract the whole application/tier/service as a single queue with one server. \cite{Kamra2004}, \cite{Villela2007}, \cite{Sharma2012}, \cite{Gandhi,Gandhi2014}, and \cite{Gergin} employed this method. Some described the cluster using a queue with multiple servers, like \cite{Ali-Eldin2012}, \cite{Jiang}, \cite{Aniello2014}, and \cite{Han2014}. Other works modeled each server as a separate queue, such as \cite{Doyle2003}, \cite{Urgaonkar2008}, \cite{Roy}, \cite{Ghanbari2012}, \cite{Kaur2014}, \cite{Spinner2014}, and \cite{Jianga}. \cite{Bi2010} proposed a hybrid model, in which the first tier is modeled as an M/M/c queue while other tiers are modeled as M/M/1 queues. Different from the traditional queuing theory, \cite{Salah2015} used an embedded Markov chain method to model the queuing system.

When the application involves multiple tiers or is composed of many services, single layer queuing models are insufficient. Instead, a network of queues is needed to describe the components and their relations. These models are known as queuing networks. As introduced in section \ref{subsec:multi-tier} and \ref{subsec:SOA}, to decide a number of resources in each component, there are two strategies. One is to divide the SLA into separate time portions and distribute them to each component. By this method, the queuing model for each component can be easily solved. However, it usually results in suboptimal solutions globally. Another method is to holistically provision resources to all the components to satisfy the SLA. Such method is more challenging as it is difficult and computationally heavy to find the optimal resource provisioning plan regarding a complex queuing network model.

Some models and methods have been proposed to tackle the challenge. \cite{Villela2007} described the model as an optimization problem and used three different approximations to simplify it. \cite{Bi2010} as well employed an optimization approach. \cite{Roy} and \cite{Zhang} utilized MVA (Mean Value Analysis), a widely adopted technique for computing expected queue lengths, waiting time at queuing nodes, and throughput in equilibrium for a closed queuing network, to anticipate the utilization at each tier under the particular provisioning. \cite{Han2014} adopted a greedy approach that continuously adds/removes one server to the most/least utilized tier until the estimated capacity is just enough to serve the current load.

As mentioned in Section~\ref{subsec:high_level_metrics}, some parameters in the queuing models are hard to measure directly, like service time. Therefore, the proposed auto-scalers should properly handle this issue. The detailed techniques were introduced in Section~\ref{subsec:high_level_metrics}.

\subsection{Machine Learning}
\label{subsec:machine_learning}

Machine learning techniques in resource estimation are applied to dynamically construct the model of resource consumption under a specific workload (online learning). In this way, different applications can utilize the auto-scalers without customized settings and preparations. They are also more robust to changes during production as the learning algorithm can self-adaptively adjust the model on the fly regarding any notable events. Online machine learning algorithms are often implemented as feedback controllers to realize self-adaptive evolution. Though offline learning can also be used to fulfill the task, it inevitably involves human intervention and thus loses the benefit of using machine learning. For works that use offline learning---if there exists any, we classify them into the application profiling category.

Despite their easiness of usage and flexibility, machine learning approaches suffer from a major drawback. It takes time for them to converge to a stable model and thus causes the auto-scaler to perform poorly during the active learning period. Certainly, the application performance is affected in this process. Furthermore, the time that is taken to converge is hard to predict and varies case by case and algorithm by algorithm.

Online learning used by the existing auto-scalers can be divided into two types: reinforcement learning, and regression.

\subsubsection{Reinforcement Learning}

Reinforcement learning aims to let the software system learn how to react adaptively in a particular environment to maximize its gain or reward. It is suitable to tackle automatic control problems like auto-scaling \cite{Tesauro2005,Zhu2012,Dutreilh2010,Dutreilh2011,Li2011,Barrett2013,Bu2013,Yazdanov2013,Fallah2015,Iqbal2015,GhobaeiArani2017}. For the auto-scaling problem, the learning algorithm's target is to generate a table specifying the best provision or deprovision action under each state. The learning process is similar to a trial-and-error approach. The learning algorithm chooses an individual operation and then observes the result. If result is positive, the auto-scaler will be more likely to take the same action next time when it faces a similar situation.

The most used reinforcement learning algorithm in the auto-scaling literature is Q-learning. A detailed description of the Q-learning algorithm and their variations in auto-scaling can be found in section 5.2 of the survey \cite{Lorido-Botran2014}.

\subsubsection{Regression}

Regression estimates the relationship among variables. It produces a function based on observed data and then uses it to make predictions. Under the context of resource estimation, the auto-scaler can record system utilization, application performance, and the workload for regression. As the training proceeds and more data is available, the predicted results also become more accurate. Although regression requires the user to determine the function type first, for example, whether the relationship is linear or quadratic, in the case of auto-scaling web applications, it is usually safe to assume a linear function.

\cite{Chen2008} used regression to dynamically build the CPU utilization model of Live Messenger given a number of active connections and login rate. The model is then used for resource provisioning. \cite{Bi2010} employed smoothing splines nonlinear regression to predict mean performance under a certain amount of resources. Then they calculated the variance based on the estimated mean. After that they used a local polynomial (LOESS) regression to map mean performance to variance. Through this method, they found out that higher workload results in both mean and variance of the response time to increase. To detect sudden changes, they rely on conducting a statistical hypothesis test of the residual distribution in two time frames with probably different sizes. If the test result is statistically significant, the model needs to be retrained. \cite{Padala2009} utilized auto-regressive moving-average (ARMA) to dynamically learn the relationship between resource allocation and application performance considering all resource types in all tiers. \cite{Gambi2013} proposed an auto-scaler using a Kriging model. Kriging models are spatial data interpolators akin to radial basis functions. These models extend traditional regression with stochastic Gaussian processes. The major advantage of them is that they can converge quickly using fewer data samples. \cite{Grimaldi2015} proposed a Proportional-Integral-Derivative (PID) controller that automatically tunes parameters to minimize integral squared error (ISE) based on a sequential quadratic programming model.

\cite{Yanggratoke2015} proposed a hybrid approach using both offline learning and online learning. They first used a random forest model and traces from a testbed to train the baseline. Then they applied regression-based online learning to train the model for real-time resource estimation.

\subsection{Hybrid Approaches}
\label{subsec:hybrid_approaches}

All the previous listed approaches have their pros and cons. Therefore, some works have integrated multiple methods together to perform resource estimation. We classify them as hybrid approaches and individually introduce them and the rationale behind such integration.

Rule-based approaches are inflexible when significant changes occur to applications and often require expert knowledge to design and test. However, if the rules can be constructed dynamically and adaptively by some learning techniques, such concern vanishes. \cite{Jing2007} and \cite{Jamshidi2016} proposed approaches that combine machine learning and fuzzy rule-based inference. They utilized machine learning to dynamically construct and adjust the rules in their fuzzy inference engine. \cite{Lamaa} first proposed a fixed fuzzy-based auto-scaler with a self-adaptive component that dynamically tunes the output scaling factor. After that, they built another fuzzy inference system as a four-layer neural network \cite{Lama2010} in which membership functions and rules can self-evolve as the time passes.

Some analytical queuing models require the observation of volatile metrics that are hard to measure directly. In these cases, a widely-adopted solution is to use machine learning approaches to estimate the concealed metrics dynamically. \cite{Gandhi} adopted Kalman filter to assess the average service time, background utilization, and end-to-end network latency. \cite{Zhang} employed application profiling and regression to learn the relationship of average CPU utilization and average service time at each tier under given request mix to solve their queuing network model using Mean Value Analysis.

To mitigate the drawback of machine learning approaches, which are slow to converge and may cause plenty of SLA violations, another model can be used to substitute the learning model temporarily and then shift it back after the learning process has converged. \cite{Tesauro2007} and \cite{Gambi2015} proposed this kind of approaches. Both of them utilized an analytical queuing model for temporary resource estimation during the training period. \cite{Tesauro2007} employed reinforcement learning while \cite{Gambi2015} adopted a Kriging-based controller for training.

To make the model more general-purpose, \cite{Chen2016} first applied feature selection and then utilized the selected features to train different models using various machine learning algorithms. The most accurate model is selected at runtime as the model for resource provisioning.

\section{Oscillation Mitigation}
Oscillation is the situation that an auto-scaler continuously performs opposite scaling operations, such as provisioning 2 VMs and then in short time deprovisioning 2 VMs. It happens when monitoring and scaling operations are too frequent, or the auto-scaler is poorly configured. Such concerns are magnified when dealing with rule-based auto-scalers whose resource estimations are relatively empirical and coarse-grained. If the scaling thresholds are poorly configured, oscillation is likely to happen. For example, suppose the scale out threshold is set to 70\%, the scale in threshold is set to 50\%, and the current utilization is 71\% with only one instance running, the auto-scaler will add one more instance to the cluster to reduce the utilization. It then quickly drops to 35\%, which is below the scale-down threshold, thus causing oscillation.

\subsection{Cooling Time}
One common solution adopted in practice \cite{Amazon2016} to mitigate oscillation is to conservatively wait a fixed minimum amount of time between each scaling operations. The time is set by users and is known as cooling time. It should be set to at least the time taken to acquire, boot up, and configure the VM. Such method is simple but effective to avoid frequent scaling operations. However, setting a long cooling time will also result in more SLA violations as the application cannot be scaled out as quickly as before. Besides, it cannot handle the situation that the auto-scaler is poorly configured.

Another way of setting the cooling time is to refine the scaling condition. Suppose the monitoring interval of the auto-scaler is 1 minute, we can achieve a prolonged scaling interval by setting the scaling trigger to how many times the monitored value exceeds the defined threshold consecutively.

\subsection{Dynamic Parameters}
Besides static cooling time, approaches that dynamically adjust parameters to reduce the possibility of causing oscillation have been proposed.

\cite{Lim2009,Lim2010} described an approach through dynamically tuning triggering thresholds for scale-down operations. The core idea is to increase the scale-down threshold when more resources are allocated to decrease the target utilization range and vice versa when resources are deallocated, which can effectively mitigate oscillation if the application resource requirement varies significantly during peak time and non-peak time. Usually, during non-peak time, a large target range is desirable to avoid the situation described in the poorly configured example, while during peak hours, a small target range is preferred to keep the utilization as close to the scale out threshold as possible.

\cite{Bodik2009} introduced a mechanism that they call, ``hysteresis parameters", to reduce oscillation. These parameters control how quickly the controller provisions and deprovisions resources. They are determined by simulations using Pegasus, an algorithm that compares different control settings to search the suitable one. \cite{Padala2009} used a stability factor to adjust the aggressiveness of the approach. As the factor increases, the control objective will be more affected by the previous allocation. As a result, the auto-scaler responds more slowly to the resulting errors caused by the previous actions in the following resource scaling windows and thus reduces oscillations. \cite{Lamaa} employed a similar approach on their fuzzy-based auto-scaler. Their approach is more advanced and flexible as the factor is self-tunable during runtime according to the resulted errors.

\cite{Ali-Eldin2012,Ali-Eldin2012a} utilized a parameter to dynamically tune the prediction horizon of the workload prediction algorithm according to its real-time error-rate. In this way, the auto-scaler can reduce the number of scaling oscillations caused by inaccurate workload predictions.

\subsection{Exhaustion}
The above methods are only capable of mitigating the possibility of oscillations. If in theory we can identify the settings that might cause oscillations and thus pose restrictions on such settings, the risk of oscillation will be eliminated. \cite{Cunha2014} and \cite{Netto2014} adopted this approach and presented models that identify the potential oscillation conditions in their rule-based auto-scalers.

\section{Scaling Timing}

When to scale the application is a critical question needed to be answered by auto-scalers. However, there is no perfect solution for this issue as different applications have diverse workload characteristics, and preference of cost and QoS. Auto-scalers can be classified into two groups based on this criterion: approaches that reactively scale the application only when necessary according to the current status of the application and the workload, and approaches that support proactive provisioning or deprovisioning of resources considering the future needs of the application.

For applications with gradual and smooth workload changes, reactive approaches are usually preferred because they can save more resources without causing a significant amount of SLA violations. In contrast, applications with drastic workload changes or strict SLA requirements often require proactive scaling before the workload increases to avoid incurring a significant amount of SLA violations during the provisioning time. Such strategy relies on prediction techniques to timely foresee incoming workload changes. Prediction is the process of learning relevant knowledge from the history and then apply it to forecast the future behaviors of some object. The assumption that behaviors are predictable lies that they are not completely random and follow some rules. Therefore, workload prediction is only viable for the workloads with patterns and thus, cannot handle the random bursts of requests, which are common in some applications, like news feed and social network. For these bursty workload scenarios, currently there is no effective solution, and we can only deal with them reactively in the best effort. Hence, regardless existence of support for proactive scaling, a qualified auto-scaler should always be able to scale reactively.

\subsection{Proactive Scaling}
As the accuracy of the prediction algorithm determines the capability of the auto-scaler to scale applications proactively, in this section, we survey prediction algorithms that have been employed by state-of-the-art works.

\subsubsection{Workload Prediction Data Source}
It is necessary to study the past workload history to understand the characteristics of the workload, including its intensity and mix during each time frame, to predict it.

Besides workload history, individual applications can rely on available information from other aspects to predict request bursts that are impossible to be derived from past workload data alone, such as weather information for an outdoor application, and political events for a news feed application. However, the relevant parameters are application-specific and thus this feature is hard to be integrated into a general purpose auto-scaler. Besides, it is also challenging to devise a prediction algorithm with real-time accuracy for resource provisioning, because there are too many parameters in the model and errors can quickly accumulate. The work by \cite{Frey2014} considers multiple outside parameters in an auto-scaler. Their approach integrates all the prediction information into a fuzzy controller.

Though it is challenging to predict the exact workload intensity using outside information, it is viable to timely detect events that may affect incoming workload intensity through social media and other channels \cite{You2013}. Since this is a broad topic itself, we focus on prediction algorithms only based on workload history.  

\subsubsection{Prediction Horizon and Control}
Typically, a prediction algorithm loops in a specified interval to predict the average or maximum workloads arriving at the application during each of the next few intervals, which form the prediction horizon. It determines how far in the future the auto-scaler aims to predict.

There are two approaches in which auto-scalers can apply prediction results in resource provisioning. The first way, which is adopted by the majority of works, takes the prediction horizon as the control interval and scales the application only based on the predicted workload of the next horizon. The other strategy is called Model Predictive Control (MPC). It sets the control interval the same to the prediction interval. When making decisions, it considers all the intervals within the horizon and determines the scaling operations at each interval using optimization. However, when executing the scaling operations, it only performs the action for the next interval and discards operations for the other intervals in the horizon. This method mitigates the problem of provisioning for short-term benefits, but it requires solving complex optimization models, and thus, consumes much more computing power. \cite{Ghanbari2012,Ghanbari2014}, and \cite{Zhang2013} employed this approach.

To tune the length of the horizon, users can either adjust the duration of each interval or number of intervals in the horizon. The size of the interval is critical to prediction precision. A large interval can significantly degrade the prediction accuracy and is useless for real-time control if the interval is greater than the control interval of the auto-scaler. The number of intervals in the horizon is also a crucial parameter, especially for the MPC approach. A balanced number should be chosen for the auto-scaler to reach good performance. If it is too small, MPC cannot fully realize its potential to make decisions for the long-term benefit. A large number, on the other hand, may mislead the auto-scaler as predictions for the intervals far in the future become increasingly inaccurate.

\subsubsection{Workload Prediction Algorithms}
Regarding workload prediction algorithms, they can be coarsely classified in two types: prediction according to recent trends and prediction based on regular patterns.

Prediction according to recent trends aims to use the workload data monitored in the near past to determine whether the workload is increasing or decreasing and how much it will change. In this case, only a few data is stored for prediction purpose. Well-established time-series analysis algorithms are commonly applied to this type of prediction tasks, such as linear regression \cite{Bodik2009,GhobaeiArani2017}, various autoregressive models (AR) \cite{Chen2008,Roy,Fang2012,Yazdanov2013,Yang2014}, and neural network-based approaches \cite{Prodan2009,Aniello2014,Nikravesh2015}. Besides using time-series analysis, \cite{Nguyen2013} proposed another method, which considers each time interval as a wavelet-based signal and then applies signal prediction techniques.

Prediction algorithms based on regular patterns assume the workload is periodic, which is valid for many applications as they tend to be more accessed during the daytime, weekdays, or the particular days in a year (tax report period, Christmas holidays). By finding these patterns, predictions can be easily made. Different from prediction algorithms based on recent trends, this type of algorithm requires a large workload archive across an extended period. Various approaches have been explored to identify workload patterns when building auto-scalers. \cite{Fang2012} employed signal processing techniques to discover the lowest dominating frequency (which corresponds to the longest repeating pattern). \cite{SilvaDias2014} utilized Holt-Winter model, which aims to identify seasonality in the workload for prediction. \cite{Jiang} devised an approach by first identifying the top K most relevant monitored data using an auto-correlation function and then employing linear regression on the selected data for prediction. \cite{Urgaonkar2008} adopted an algorithm based on the histogram for the workload with daily patterns.

\cite{Herbst2014} integrated many predictors into one auto-scaler. They presented an approach to dynamically select appropriate prediction methods according to the extracted workload intensity behavior (WIB, simply the workload characteristics) and user's objectives. The mappings of prediction methods to WIBs are stored in a decision tree and are updated during runtime based on the recent accuracy of each algorithm. 

\subsubsection{Resource Usage Prediction}
Instead of predicting workload, it is also possible to directly predict resulted resource usage according to the historical usage data. This strategy is commonly used by auto-scalers that only support vertical scaling, as for a single machine, resource usage can substitute workload intensity. Some proposals \cite{Islam2010,Caron2011,AlmeidaMorais2013} that target horizontal scaling also follow this strategy to accomplish both workload prediction and resource estimation together.

\cite{Gong2010} used signal processing to discover the longest repeating pattern of resource usage and then relied on dynamic time warping (DTW) algorithm to make the prediction. For applications without repeating patterns, they referred to a discrete-time Markov chain with finite states to derive a near prediction of future values. \cite{Islam2010} explored using linear regression and neural network to predict CPU usage. \cite{Caron2011} adopted a pattern matching approach which abstracts it as a string matching problem and solved it using the Knuth-Morris-Pratt (KMP) algorithm. \cite{Yazdanov2012} utilized an auto-regressive (AR) method to predict short-term CPU usage. \cite{AlmeidaMorais2013} employed multiple time series algorithms to predict CPU usage, and based on their runtime accuracy, the best is selected. \cite{Loff2014} also used various prediction algorithms. However, instead of selecting the best one, their approach combines the results of different predictors using weighted k-Nearest Neighbors algorithm. The weight of each predictor is dynamically adjusted according to their recent accuracy.

\section{Scaling Methods}
Depending on the particular cloud environment, elastic scaling can be performed vertically, horizontally, or both. Each of them has their advantages and limitations. In this section, we discuss the key factors that need to be considered when making the provisioning plan.

\subsection{Vertical Scaling --- VM Resizing}
Vertical scaling means removing or adding resources, including CPU, memory, I/O, and network, to or from existing VMs. To dynamically perform these operations, modern hypervisors utilize mechanisms such as CPU sharing and memory ballooning, to support CPU and memory hot-plug. However, major cloud providers, such as Amazon, Google, and Microsoft, do not support adjusting resources during runtime. In these platforms, it is essential to shut down the instance first to add resources. Some providers like Centurylink\footnote{https://www.ctl.io/autoscale/} allow users to scale CPU cores without downtime vertically. Profitbricks\footnote{https://www.profitbricks.com/help/Live\_Vertical\_Scaling} permits to add both CPU and memory to the VMs dynamically.

Vertical scaling is considered not suitable for highly scalable applications due to its limitations. Ideally, the maximum capacity a VM can scale to is the size of the physical host. However, multiple VMs are usually residing on the same physical machine competing for resources, which further confines the potential scaling capability. Though limited, dynamic vertical scaling outperforms horizontal scaling in provisioning time as it can be in effect instantaneously. Besides, some services or components that are difficult to replicate during runtime, such as database server, and stateful application servers, can be benefited by vertical scaling. \cite{Dawoud2012} conducted an experimental study of vertical scaling using RUBBOS benchmark on both its application server and database, which highlights the advantages of vertical scaling mentioned above.

Many auto-scalers have been developed using solely vertical scaling to manage VMs on the same physical host. Some of them only considered scaling CPU resources \cite{Kalyvianaki2009,Shen2011,Yazdanov2012,Spinner2014}. Some targeted both CPU and memory \cite{Gong2010,Zhu2012,Dawoud2012,Yazdanov2013,Chen2015,Wang2016}. \cite{Jing2007} focused on CPU in the prototype and claimed their method could be extended to other resources. \cite{Bu2013} proposed an approach that adjusts not only CPU and memory allocation but also application parameters. \cite{Padala2009} scaled both CPU and disk. These auto-scalers are experimental prototypes designed to be deployed in either private or public clouds. However, they have not been included in any commercial offerings yet.

\subsection{Horizontal Scaling --- Launching New VMs}

Horizontal scaling is the core of the elasticity feature of the cloud. Most cloud providers offer standardized VMs of various sizes for customers to choose. Others allow users to customize their VMs with a specific amount of cores, memory, and network bandwidth. Besides, multiple pricing models co-exist in the current cloud market, which further increases the complexity of the provisioning problem.

\subsubsection{Heterogeneity}
Regarding a single tier/service within a web application, if the billing is constant, the use of homogeneous VMs is well acceptable as it is easy to manage. The auto-scaling services offered by cloud providers only allow the use of homogeneous VMs. Selecting which type of VM to use is considered the responsibility of users in commercial auto-scalers. The optimal solution depends on the resource profile of the tier/service, e.g., whether it is CPU or memory intensive, and the workload characteristic. Under large workloads, using small or large instance results in little difference as the remainder of resources that are not used accounts very small percentage of the total resources. While under a small and fluctant workload smaller instances are preferred, as scaling can be conducted in finer granularity and thus saves cost.

Cost-efficiency of VMs is highly co-related to the application and workload. If changes happen to them, the choice of VM type should also be reconfigured. \cite{Grozev2016} proposed a method that detects changes online using the Hierarchical Temporal Memory (HTM) model and a dynamically trained artificial neural network (ANN) and then reselects the most cost-efficient VM type.

The use of heterogeneous VMs to scale web applications has been explored in the literature. Under conventional billing, where price grows linearly with VM's capability, heterogeneity can bring some extra cost-saving but not significant. Furthermore, it the search of the optimal provisioning plan with a combination of heterogeneous VMs is often computing-intensive. \cite{Srirama2014} employed linear programming to solve the provisioning problem, yet only achieved limited cost saving against AWS auto-scaling. \cite{Fernandez2014} abstracted the provisioning combinations as a tree and searched the proper provisioning by traversing the tree according to different SLAs. \cite{Sharma2012} used a greedy algorithm to obtain the provisioning plan. In a different scenario in which the capability of VMs increases exponentially to their prices, heterogeneity has the potential to save significant cost, which is shown in \cite{Sedaghat2013} and \cite{Upendra}. They both consider the transition cost (the time and money spent to convert from the current provisioning to the target provisioning) and the cost of resource combination in the optimization problem.

We proposed an auto-scaler \cite{Qu2016} that uses heterogeneous spot instances to provision web applications. The intention of using heterogeneous VMs in this case is to boost the reliability of clusters based on spot instances to save cost, which is explained in the following section.

\subsubsection{Pricing Models}

The current cloud pricing models can be classified into three types by pricing model: on-demand, reserved, and rebated. In on-demand mode, the provider sets a fixed unit price for each type of VM or unit of certain resource, and charges the user by units of consumption. Users are guaranteed to obtain the required resources and agreed performance, which most auto-scalers assume the target application is adopting. The reserved mode requires the user to pay an upfront fee for cheaper use of a certain amount of resources. If highly utilized, users can save a considerable sum of money than acquiring resources in on-demand mode. Providers create the rebate mode aiming to sell their spare capacity. They are usually significantly cheaper than on-demand resources. There are several ways to offer rebated resources. Amazon employed an auction-like mechanism to sell instances, called spot instances. In this mode, the user is required to submit a bid on the resources. Suppose the bid exceeds the current market price, the bid is fulfilled and the user is only charged for the current market price. The acquired spot instances are guaranteed to have the same performance of their on-demand counterparts. However, they are reclaimed whenever the market price goes beyond user's bidding price. Google offers their spare capacity as preemptible VMs. Different from Amazon, they set a fixed price to the VM, which is 30\% of the regular price, and the VM is available at most for 24 hours. Rebated instances are considered not suitable to host web applications that are availability-critical as they can be reclaimed by providers at any time.

ClusterK\footnote{http://www.geekwire.com/2015/amazon-buys-clusterk-a-startup-that-lets-developers-run-aws-workloads-more-cheaply/ acquired by AWS in 2015}, and our previous work \cite{Qu2016}  demonstrated that it is feasible to build an auto-scaler utilizing spot instances by exploiting various market behaviours of different spot markets to achieve both high availability and considerable cost saving. \cite{Sharma2015,He2015} endeavored to build a reliable general-purpose cloud platform upon spot instances using nested virtualization. As these systems lie in the underlying virtualization layer, all types of auto-scaling techniques can be applied to them.

Pricing models can also be classified according to their billing period, which is the minimum unit consumption. Providers have set their billing period to every minute, hour, day, week, month, or year. The length of the billing period has a significant impact on the cost-efficiency for elasticity. Obviously, the shorter the billing period, the more flexible and cost-efficient it is for auto-scaling. If the billing period exceeds the order of hour, there is no use in applying auto-scaling to save resource cost, since provisioning for the peak load of the day incurs the same cost.

\subsection{Hybrid}

As mentioned previously, horizontal scaling is slow in provisioning and vertical scaling is restricted by resources available in the host. It is natural to employ vertical scaling and horizontal scaling together to mitigate these issues. The idea is to utilize vertical scaling when possible to quickly adapt to changes and only conduct horizontal scaling when vertical scaling reaches its limit. \cite{Urgaonkar2008}, \cite{Huber2011}, \cite{Rui2012}, and \cite{Yang2014} followed this strategy. 

Mixing vertical scaling and horizontal scaling can also bring cost benefit. \cite{Dutta}, and \cite{Gandhi2014} explored optimization techniques to search for the scaling plan that incurs the least cost with a hybrid of vertical and horizontal scaling.

Vertical scaling and horizontal scaling can be separately applied to different components of the application as well since some parts such as database servers are difficult to be horizontally scaled. \cite{Nisar} demonstrated this approach in a case study.

\section{Environment}
\label{sec:detail_end}

In the previous sections, we discussed the characteristics of auto-scalers operating in a single cloud data center. In this section, we first give a summary of single cloud auto-scalers. After that, we introduce their counterparts that can coordinately work across multiple cloud data centers.

\begin{landscape}
\scriptsize
\begin{longtable} {c | c | c | c | c | c | c | c | c}\caption{A Review of auto-scaling properties of key works for single cloud}\label{tab:auto_scaling_review}\\
\hline
\hline
\multicolumn{1}{c |}{\textbf{Work}} & \multicolumn{1}{m{1.7cm} |}{\textbf{Application Architecture}} & \multicolumn{1}{m{1cm} |}{\textbf{Sticky Session}} & \multicolumn{1}{c |}{\textbf{Adaptivity}} & \multicolumn{1}{m{1.35cm} |}{\textbf{Scaling Indicators}} & \multicolumn{1}{c |}{\textbf{Resource Estimation}} & \multicolumn{1}{m{1.5cm} |}{\textbf{Oscillation Mitigation}} & \multicolumn{1}{c |}{\textbf{Proactive}} & \multicolumn{1}{c}{\textbf{Scaling Methods}}\\
\endfirsthead

\multicolumn{9}{c}
{{\bfseries \tablename\ \thetable{} -- continued from previous page}} \\
\hline
\hline
\multicolumn{1}{c |}{\textbf{Work}} & \multicolumn{1}{m{1.7cm} |}{\textbf{Application Architecture}} & \multicolumn{1}{m{1cm} |}{\textbf{Sticky Session}} & \multicolumn{1}{c |}{\textbf{Adaptivity}} & \multicolumn{1}{m{1.35cm} |}{\textbf{Scaling Indicators}} & \multicolumn{1}{c |}{\textbf{Resource Estimation}} & \multicolumn{1}{m{1.5cm} |}{\textbf{Oscillation Mitigation}} & \multicolumn{1}{c |}{\textbf{Proactive}} & \multicolumn{1}{c}{\textbf{Scaling Methods}}\\
\endhead

\multicolumn{9}{r}{{Continued on next page}} \\ 
\endfoot

\endlastfoot

\hline
\cite{Doyle2003} & single-tier & \Checkmark & non-adaptive & hybrid & analytical model & --- & \XSolid & vertical\\
\hline
\cite{Kamra2004} & 3-tier & \Checkmark & self-adaptive & high-level & analytical model & --- & \XSolid & vertical\\
\hline
\cite{Tesauro2005} & single-tier & \XSolid & self-adaptive & high-level & reinforcement learning & --- & \XSolid & hom. horizontal\\
\hline
\cite{Tesauro2007} & single-tier & \XSolid & self-adaptive & high-level & hybrid & --- & \XSolid & hom. horizontal\\
\hline
\cite{Jing2007} & single-tier & \Checkmark & self-adaptive & high-level & hybrid & --- & \XSolid & vertical\\
\hline
\cite{Villela2007} & single-tier & \XSolid & non-adaptive & hybrid & analytical model & --- & \XSolid & hom. horizontal\\
\hline
\cite{Zhang} & multi-tier & --- & --- & hybrid & hybrid & --- & --- & --- \\
\hline
\cite{Chen2008} & single-tier & \Checkmark & self-adaptive & hybrid & regression & --- & \Checkmark & hom. horizontal\\
\hline
\cite{Urgaonkar2008} & multi-tier & \XSolid & non-adaptive & high-level & analytical model & --- & \Checkmark & hybrid\\
\hline
\cite{Iqbal2009} & single-tier & \XSolid & non-adaptive & high-level & rule-based & --- & \XSolid & hetr. horizontal\\
\hline
\cite{Lim2009} & single-tier & \XSolid & self-adaptive & low-level & rule-based & dynamic para. & \XSolid & hom. horizontal\\
\hline
\cite{Bodik2009} & single-tier & \XSolid & self-adaptive & high-level & regression & dynamic para. & \Checkmark & hom. horizontal\\
\hline
\cite{Padala2009} & multi-tier & \Checkmark & self-adaptive & hybrid & regression & dynamic para. & \XSolid & vertical\\
\hline
\cite{Kalyvianaki2009} & single-tier & \Checkmark & self-adaptive & low-level & rule-based & --- & \XSolid & vertical\\
\hline
\cite{Lamaa} & multi-tier & \XSolid & self-adaptive & high level & fuzzy inference & dynamic para. & \XSolid & hom. horizontal\\
\hline
\cite{Lim2010} & storage-tier & \XSolid & self-adaptive & low-level & rule-based & dynamic para. & \XSolid & hom. horizontal\\
\hline
\cite{Dutreilh2010} & single-tier & \XSolid & self-adaptive & high-level & hybrid & cooling time & \XSolid & hom. horizontal\\
\hline
\cite{Gong2010} & single-tier & \Checkmark & self-adaptive & low-level & hybrid & --- & \Checkmark & vertical\\
\hline
\cite{Islam2010} & single-tier & \XSolid & self-adaptive & low-level & neural net./regression & --- & \Checkmark & ---\\
\hline  
\cite{Lama2010} & multi-tier & \XSolid & self-adaptive & high-level & hybrid & --- & \XSolid & hom. horizontal\\
\hline
\cite{Bi2010} & multi-tier & \XSolid & non-adaptive & high-level & analytical model & --- & \XSolid & hom. horizontal\\
\hline
\cite{Singh2010} & multi-tier & \XSolid & non-adaptive & high-level & analytical model & --- & \XSolid & hom. horizontal\\
\hline
\cite{Jianga} & SOA & \XSolid & self-adaptive & high-level & analytical model & --- & \XSolid & hom. horizontal\\
\hline 
\cite{Chieua} & single-tier & \Checkmark & non-adaptive & high-level & rule-based & --- & \XSolid & hom. horizontal\\
\hline
\cite{Dutreilh2011} & single-tier & \XSolid & self-adaptive & high-level & reinforcement learning & --- & \XSolid & hom. horizontal\\
\hline
\cite{Li2011} & single-tier & \XSolid & self-adaptive & low-level & reinforcement learning & --- & \XSolid & hom. horizontal\\
\hline
\cite{Caron2011} & single-tier & \XSolid & self-adaptive & low-level & string matching & --- & \Checkmark & ---\\
\hline
\cite{Huber2011} & single-tier & \XSolid & non-adaptive & hybrid & rule-based & --- & \XSolid & hybrid\\
\hline
\cite{Iqbal2011} & multi-tier & \XSolid & self-adaptive & hybrid & hybrid & --- & \XSolid & hom. horizontal\\
\hline
\cite{Jiang2011} & multi-tier & \XSolid & non-adaptive & hybrid & online profiling & --- & \XSolid & hetr. horizontal\\
\hline
\cite{Malkowski2011} & multi-tier & \XSolid & self-adaptive & hybrid & hybrid & --- & \XSolid & hom. horizontal\\
\hline
\cite{Roy} & multi-tier & \XSolid & non-adaptive & hybrid & analytical model & --- & \Checkmark & hom. horizontal\\
\hline
\cite{Upendra} & multi-tier & \XSolid & non-adaptive & high-level & profiling & --- & \Checkmark & hetr. horizontal\\
\hline
\cite{Vasic2012} & single-tier & \XSolid & self-adaptive & low-level & online profiling & --- & \XSolid & hom. horizontal\\
\hline
\cite{Ali-Eldin2012} & single-tier & \XSolid & self-adaptive & high-level & analytical model & dynamic para. & \Checkmark & hom. horizontal\\
\hline
\cite{Ali-Eldin2012a} & single-tier & \XSolid & self-adaptive & high-level & analytical model & dynamic para. & \Checkmark & hom. horizontal\\
\hline
\cite{Dawoud2012} & single-tier & \XSolid & non-adaptive & low-level & rule-based & --- & \XSolid & compare ver. hor.\\
\hline
\cite{Fang2012} & single-tier & \XSolid & --- & --- & --- & --- & \Checkmark & ---\\
\hline
\cite{Yazdanov2012} & single-tier & \Checkmark & self-adaptive & low-level & regression & --- & \Checkmark & vertical\\
\hline
\cite{Ghanbari2012} & single-tier & \XSolid & non-adaptive & high-level & analytical model & --- & \XSolid & hetr. horizontal\\
\hline
\cite{Zhu2012} & single-tier & \Checkmark & self-adaptive & low-level & reinforcement learning & --- & \XSolid & vertical\\
\hline
\cite{Dutta} & multi-tier & \XSolid & non-adaptive & hybrid & application profiling & --- & \XSolid & hybrid\\
\hline
\cite{Gandhi2012} & multi-tier & \XSolid & non-adaptive & hybrid & profiling & --- & \XSolid & hom. horizontal\\
\hline 
\cite{Rui2012} & multi-tier & \XSolid & non-adaptive & high-level & rule-based & --- & \XSolid & hybrid\\
\hline
\cite{Sharma2012} & multi-tier & \XSolid & non-adaptive & high-level & analytical model & --- & \XSolid & hetr. horizontal\\
\hline
\cite{Jiang} & single-tier & \XSolid & non-adaptive & high-level & analytical model & --- & \Checkmark & hom. horizontal\\
\hline
\cite{Al-Haidari2013} & single-tier & \XSolid & non-adaptive & high-level & rule-based & --- & \XSolid & hom. horizontal\\
\hline
\cite{Bu2013} & single-tier & \Checkmark & self-adaptive & high-level & reinforcement learning & --- &  \XSolid & vertical\\
\hline
\cite{Gambi2013} & single-tier & \XSolid & self-adaptive & low-level & Kriging regression & --- & \XSolid & hom. horizontal\\
\hline
\cite{Barrett2013} & single-tier & \XSolid & self-adaptive & high-level & reinforcement learning & --- & \XSolid & hetr. horizontal\\
\hline
\cite{Sedaghat2013} & single-tier & \XSolid & non-adaptive & high-level & --- & --- & \XSolid & hetr. horizontal\\
\hline
\cite{Yazdanov2013} & single-tier & \XSolid & self-adaptive & hybrid & reinforcement learning & --- & \Checkmark & vertical\\
\hline
\cite{Ali-Eldin2013} & single-tier & \XSolid & switch & --- & --- & --- & \Checkmark & hom. horizontal\\
\hline 
\cite{AlmeidaMorais2013} & single-tier & \XSolid & self-adaptive & low-level & various regressions & --- & \Checkmark & hom. horizontal\\
\hline
\cite{Nguyen2013} & multi-tier & \XSolid & non-adaptive & hybrid & online profiling & --- & \Checkmark & hom. horizontal\\
\hline
\cite{Herbst2014} & --- & \XSolid & self-adaptive & --- & --- & --- & \Checkmark & ---\\
\hline
\cite{Grozev2014} & single-tier & \Checkmark & non-adaptive & low-level & rule-based & --- & \XSolid & hom. horizontal\\
\hline
\cite{SilvaDias2014} & single-tier & \XSolid & non-adaptive & hybrid & rule-based & --- & \Checkmark & hom. horizontal\\
\hline
\cite{Loff2014} & single-tier & \XSolid & non-adaptive & low-level & rule-based & --- & \Checkmark & hom. horizontal\\
\hline
\cite{Cunha2014} & single-tier & \XSolid & self-adaptive & low-level & rule-based & theory & \XSolid & hom. horizontal\\
\hline
\cite{Netto2014} & single-tier & \XSolid & self-adaptive & low-level & rule-based & theory & \XSolid & hom. horizontal\\
\hline
\cite{Aniello2014} & single-tier & \XSolid & non-adaptive & high-level & analytical model & --- & \Checkmark & hom. horizontal\\
\hline
\cite{Frey2014} & single-tier & \XSolid & non-adaptive & hybrid & fuzzy inference & --- & \Checkmark & hom. horizontal\\
\hline
\cite{Yang2014} & single-tier & \XSolid & non-adaptive & hybrid & rule-based & --- & \Checkmark & hetr. horizontal\\
\hline
\cite{Fernandez2014} & single-tier & \XSolid & non-adaptive & high-level & profiling & --- & \XSolid & hetr. horizontal\\
\hline
\cite{Srirama2014} & single-tier & \XSolid & non-adaptive & --- & --- & --- & \XSolid & hetr. horizontal\\
\hline
\cite{Gandhi2014} & single-tier & \XSolid & self-adaptive & high-level & analytical model & --- & \XSolid & hybrid\\
\hline
\cite{Spinner2014} & single-tier & \Checkmark & self-adaptive & hybrid & analytical model & --- & \XSolid & vertical\\
\hline
\cite{Gergin} & multi-tier & \XSolid & non-adaptive & high-level & analytical model & --- & \XSolid & hom. horizontal\\
\hline
\cite{Han2014} & multi-tier & \XSolid & non-adaptive & high-level & analytical model & --- & \XSolid & hom. horizontal\\
\hline
\cite{Kaur2014} & multi-tier & \XSolid & non-adaptive & high-level & analytical model & --- & \Checkmark & hom. horizontal\\
\hline
\cite{Gandhi} & multi-tier & \XSolid & self-adaptive & hybrid & analytical model & --- & \XSolid & hom. horizontal\\
\hline
\cite{Nikravesh2015} & --- & --- & --- & --- & --- & --- &  \Checkmark & ---\\
\hline
\cite{Yanggratoke2015} & single-tier & \XSolid & self-adaptive & high-level & batch \& online learning & --- & \XSolid & hom. horizontal\\
\hline
\cite{Grimaldi2015} & single-tier & \XSolid & self-adaptive & low-level & rule-based & --- & \XSolid & hom. horizontal\\
\hline
\cite{Gambi2015} & single-tier & \XSolid & self-adaptive & high-level & hybrid & --- & \XSolid & hom. horizontal\\
\hline
\cite{Salah2015} & single-tier & \XSolid & non-adaptive & high-level & analytical model & --- & \XSolid & hom. horizontal\\
\hline
\cite{Iqbal2015} & multi-tier & \XSolid & self-adaptive & high-level & reinforcement learning & --- & \XSolid & hom. horizontal\\
\hline
\cite{Chen2015} & single-tier & \Checkmark & self-adaptive & hybrid & analytical model & --- & \XSolid & vertical\\
\hline
\cite{Amazon2016} & single-tier & \XSolid & non-adaptive & high/low & rule-based & cooling time & \XSolid & hom. horizontal\\
\hline
\cite{RightScale2016} & single-tier & \XSolid & non-adaptive & high/low & rule-based & cooling time & \XSolid & hom. horizontal\\
\hline
\cite{Qu2016} & single-tier & \XSolid & non-adaptive & low-level & profiling & --- & \XSolid & hetr. horizontal\\
\hline
\cite{Jamshidi2016} & single-tier & \XSolid & self-adaptive & high-level & hybrid & --- & \XSolid & hom. horizontal\\
\hline
\cite{Grozev2016} & single-tier & \XSolid & self-adaptive & hybrid & rule-based & --- & \XSolid & hetr. horizontal\\
\hline
\cite{Chen2016} & single-tier & \XSolid & self-adaptive & hybrid & hybrid & --- & --- & ---\\
\hline
\cite{Wang2016} & single-tier & \Checkmark & self-adaptive & hybrid & analytical modeling & --- & \Checkmark & vertical\\
\hline
\cite{GhobaeiArani2017} & single-tier &\XSolid & self-adaptive & hybrid & reinforcement learning & --- & \Checkmark & hom. horizontal\\
\hline
\hline

\end{longtable}
\end{landscape}

\subsection{Single Cloud}
The auto-scaling challenges and developments in single cloud environments have been thoroughly covered in the previous sections. The auto-scaling process is abstracted as an MAPE loop, and within each phase of the loop, we have identified corresponding design challenges. Based on the taxonomy and explanation of the concepts, we map each work to the specific categories for each discussed feature in Table \ref{tab:auto_scaling_review}. Readers can refer to it to quickly grasp the general design of each surveyed auto-scaler.

\subsection{Multiple Clouds}

Modern applications are often deployed in multiple cloud data centers for various purposes \cite{Grozev2014a}: 1) multi-cloud deployment helps reducing response latency if users can be served by the nearest data center; 2) it improves availability and reliability of the application against data center outages by replicating the application stack in multiple regions; 3) it enables the service provider to exploit cost differences among different vendors; and 4) it prevents vendor lock-in. Auto-scalers should be able to support this type of deployment as well.

When expanded to multiple clouds, auto-scaling remains the same problem if applications in different cloud data centers are managed completely standalone, which is the common practice of the industry. In this case, usually the service provider firstly selects a set of cloud data centers to host the application. Each data center is intended only to serve requests coming from nearby users and is separately managed by a dedicated local auto-scaler without global coordination of request routing and resource provisioning.

Though easy to manage, such strategy is not optimal in an environment where both workload and resource price are highly dynamic. As time passes, it is better to move resources to cheaper data centers to save cost, or to data centers that are closer to certain groups of users to improve their QoS. Auto-scaling becomes more complicated in these scenarios as it not only needs to make decisions on resource provisioning but also location selection and request routing.

Some works explored holistic solutions for resource management of web applications in multiple clouds. They can be further divided in two types. The first type always deploys the whole application stacks in the chosen data centers. The other type allows separate deployment of application components in different data centers.
\cite{Zhang2013} and \cite{Rodolakis2006} targeted the first type of problems. \cite{Zhang2013} assumed that each potential data center is capped, which is in contrast to the common illusion that cloud data centers have ``infinite'' amount of resources, and applications are deployed in one VM. Their objective is to minimize the total cost of resources used by applications through dynamically acquiring and releasing servers from geographically dispersed data centers under the constraint of demand, capacity, and SLA. They employed the Model Predictive Control (MPC) framework and a quadratic optimization model to adjust resource allocation in each data center and request routing from each location. Differently, \cite{Rodolakis2006} considered a scenario that without data center capacity constraints and dynamic pricing. They dissected the problem into three parts and proposed approximation algorithms for each of them to form an integrated solution. \cite{Calcavecchia2012} devised a decentralized auto-scaler for multiple clouds. It can autonomously start VMs at regions suffering from insufficient capacity through a voting mechanism. \cite{Guo2016} proposed a geo-aware auto-scaler which not only predicts the amount of workload in the future, but also the distribution of workload. By doing this, it is able to start and shut down VMs in data centers that are close to the places of workload changes.

Regarding the second problem type, \cite{Tortonesi2016} proposed a genetic-based algorithm to search the deployment of a two-tier application across multiple clouds with minimum resource and SLA violation cost. \cite{Rochman2014} modeled the problem as a min-cost flow problem and solved it with Bipartite Graph Algorithm. \cite{Grabarnik2014} added more complexity to the problem by also optimizing the chosen VM types for each tier in a multi-tier application. They devised a 2-phase metaheuristic algorithm with the outer phase responsible for assigning components to data centers also using a genetic-based algorithm, and the inner phase using a random search algorithm to map the components to specific types of VMs. None of these solutions bears reliability in mind, which is necessary for this kind of deployment. If poorly planned, instead of improving reliability and availability of the application, dispersing replicas into multiple data centers can create multiple points of failures and substantially reduce uptime. It is important for auto-scalers to ensure that every component is properly replicated in multiple data centers all the time. 

The cost of data replication is another reason makes it beneficial to provide a holistic solution for auto-scaling in multiple clouds for some applications, such as video streaming applications. For these applications, QoS cannot be met without enough bandwidth between the video storage site and the end customer. The simplest solution to the bandwidth provisioning solution is to replicate all the videos in every data center and serve each customer from the one with sufficient bandwidth available. However, it is unrealistic and extremely wasteful. The service provider needs to decide for each video how many replicas it should keep and where they should be placed to save cost. Along with the data, serving applications should be co-located as well, and user requests need to be properly diverted to particular serving replicas because of the bandwidth limit of serving VMs. To realize the above targets, \cite{Wu2012} proposed and implemented a prototype using Model Predictive Control and subgradient algorithm.

The mentioned holistic approaches require solving complex optimization problems, which takes considerable time, making them only applicable to perform auto-scaling in coarse-grained time intervals and limiting their ability to react to drastic workload changes. Therefore, for applications with highly variable workloads, the choice of using holistic approaches is doubtful. In these cases, the local auto-scalers can be deployed in each data center to handle the fine-grained scaling needs.

\section{Discussion and Future Directions}
\label{sec:discussion}

According to the taxonomy and analysis, it is clear that there are gaps between the current solutions and an ideal auto-scaler in various aspects. In the following section, we discuss them and point out potential methods and directions to improve current solutions.

\subsection{Service-based Architectures}
The research on scaling complex applications following service-based architectures is still at early stage and limited literature can be found in this area. Moreover, due to lack of accurate resource estimation models, only a simple approach that tentatively and recursively provision resources to a selected service is proposed, which takes a long time to reach the overall target performance. If accurate resource estimation model is available for service-based applications, the auto-scaler can provision resources in one shot to every service with minimum provision time. Models using queuing networks can be explored to fulfill the gap. It also calls for efficient online optimization algorithms to decide how each service should be provisioned in real-time to minimize cost.

\subsection{Monitoring Tools for Hidden Parameters}
It is important to implement low-cost monitoring tools that can provide real-time measurement of unknown parameters, such as average service time, and request mix, for general purpose applications to facilitate accurate resource estimation and provisioning. Because of the intrusive nature of these parameters, such tools can be integrated into application service containers.

\subsection{Resource Estimation Models}
Although plenty of resource estimation models have been proposed for various types of application architectures, they still need to be improved in accuracy, generality, computing requirements, and ease of use. We believe hybrid estimation models that encompass strengths of both analytical modeling and machine learning approaches are the most promising ones. Other directions, such as general purpose queuing network models for service-based applications, and efficient and accurate online profiling techniques, are important and need to be further investigated.

\subsection{Provisioning using Rebated Pricing Models}

Besides Amazon's spot cloud, providers like Google and Microsoft have introduced their rebated pricing models. However, studies have only concentrated on exploring how to utilize Amazon's spot market while have been oblivious to other providers offerings. New works can aim to use cost models from other providers to provision resources. It is also interesting to research the use of rebated resources in a multiple cloud environment with resources from multiple data centers of the same provider or from multiple providers to minimize cost under QoS constraints. Besides, the proposed approaches only combine on-demand resources with rebated resources. Auto-scalers that can employ on-demand, reserved, and rebated resources would be useful in industry, which can be another potential future research direction.

\subsection{Better Vertical Scaling Support}

Only a few providers enable users to vertically scale up their VMs without downtime, and none of them allow live scaling down, mainly due to the induced complication in the resource management of the data centers as it will result in more VM live migrations, and limitations of operating systems. More research needs to be conducted to ease providers to enable vertical scaling option in their infrastructure, which involves proposing vertical-scaling-aware VM allocation and live migration algorithms, devising and implementing generic vertical scaling APIs, and enhancing support for vertical scaling in hypervisors and operating systems.

\subsection{Event-based Workload Prediction}

As mentioned before, existing auto-scalers mostly rely on past workload history to predict future workload. With the growing popularity of social media and other real-time information channels, it is interesting to investigate the use of these sources of information to predict workload burst accurately. Although it is difficult to design a general-purpose predictor of this kind for various applications, there is potential to build auto-scalers that cater to the characteristics of a certain type of applications that can benefit from this approach, such as news applications whose workloads are boosted by events in the physical world, and outdoor applications whose workloads are subject to weather conditions.

\subsection{Reliability-aware Multi-cloud Auto-scaling}

Holistic auto-scaling solutions in multi-cloud environments ignore the impact on application availability caused by data center outages. It is necessary to address this issue before holistic approaches can be applied in a production scenario, which requires new models that quantitively measure the level of reliability for specific deployments and include reliability requirement as a constraint in the optimization problem.

\subsection{Energy and Carbon-aware Auto-scaling}

Existing works only focus on financial cost and QoS aspects. As another primary concern of the ICT sector, energy and carbon footprint should also be considered in the auto-scalers. Nowadays, many data centers are equipped with on-site generators utilizing renewable energy. However, these sources of energy, such as wind and solar, are unstable. At the cloud provider level, the auto-scalers can gather the real-time energy usage information and preferentially provision resources in data centers that have renewable energy available to maximize use of on-site renewable energy. Within a single data center, auto-scalers can utilize vertical scaling as much as possible to avoid starting new physical machines to save energy. 

\subsection{Infrastructure-level Auto-scaling with User Preferences}

From a provider's perspective, enabling auto-scaling helps them to cut cost and meet environmental obligations by reducing electricity consumption and carbon emission. However, they often do not have the freedom to allocate resources asked by users to whichever data center they own, which limits their ability to maximize their savings. For example, the cloud provider prefers to start VMs in its US data center because it generates extra solar energy; while the application provider hopes the resources can be allocated in Europe, which is the target market that the application is serving. It is an interesting problem to study that how should cloud providers automatically allocate resources for their customers to minimize their own bills while respecting each user's individual preferences regarding the time and space locality of their applications, government regulations, and resource types.

\subsection{Container-based Auto-scalers}

The emergence of containers, especially container-supported microservices and service pods, has raised a new revolution in web application resource management. However, dedicated auto-scaling solutions that cater to the specific characteristics of the container era are still left to be explored. Though this survey focuses on auto-scalers based on VMs, we believe some of the notions and techniques mentioned in this paper can inspire research of container-based auto-scalers as the core requirements of them are similar. For example, current workload prediction and oscillation mitigation techniques can be directly applied to container-based systems as they are not dependent on the underlying platforms. However, in some aspects they are different, for example, containers are more flexible in size and quicker to provision, which results in a larger optimization space for making scaling plans. Besides, containers are more susceptible to noisy neighbours as they offer weaker level of isolation than VMs. To make things worse, modern container management systems, such as Mesos, Kubernetes, and Openshift, schedule different applications on a shared cluster. These issues can cause resource contentions and high variability in processing power, which should be considered in the resource estimation process. Last but not least, the container-based auto-scaling problem is also mixed with the resource allocation problem as containers need to be efficiently scheduled and consolidated on physical hosts or VMs to save cost.

\section{Summary and Conclusions}

Auto-scaling is a technique that automatically adjusts resources provisioned to applications according to real-time workloads without human intervention. It helps application providers minimize their resource bills of using cloud resources while meeting QoS expectations of their customers. However, designing and implementing an auto-scaler faces many challenges. Many research works have targeted this problem and many auto-scalers with diverse characteristics have been proposed.

In this paper, we surveyed the developments of auto-scaling techniques for web applications in clouds. Auto-scaling can be abstracted as a MAPE (Monitoring, Analysis, Planning, and Execution) loop. We identified key challenges that need to be addressed in each phase of the loop and presented a taxonomy of auto-scalers regarding their key properties. Our taxonomy comprehensively covers the listed challenges and categorizes the works based on their solutions to each problem. According to the taxonomy, we analyzed existing techniques in detail to discuss their strength and weaknesses. Based on the analysis, we proposed promising directions that the research community can pursue in the future.

% Acknowledgments
\begin{acks}

We thank Dr. Adel Nadjaran Toosi, Dr. Amir Vahid Dastjerdi, Dr. Yaser Mansouri, Xunyun Liu, Minxian Xu, and Bowen Zhou for their valuable comments and suggestions in improving the quality of the paper.
\end{acks}

% Bibliography
\bibliographystyle{ACM-Reference-Format-Journals}
\bibliography{survey}
                             % Sample .bib file with references that match those in
                             % the 'Specifications Document (V1.5)' as well containing
                             % 'legacy' bibs and bibs with 'alternate codings'.
                             % Gerry Murray - March 2012

% History dates
%\received{February 2007}{March 2009}{June 2009}

% Electronic Appendix
%\elecappendix

%\medskip

\end{document}